\documentclass[aps,pre,twocolumn,showpacs,superscriptaddress]{revtex4}

\usepackage{graphicx}
\usepackage{dcolumn}
\usepackage{bm}
\usepackage{amssymb}

\usepackage{color}

\begin{document}

\title{Membrane simulation models from nm to $\mu$m scale}

\author{Hiroshi Noguchi}
\affiliation{Institute for Solid State Physics, University of Tokyo,
 Kashiwa, Chiba 277-8581, Japan}
\affiliation{Institut f\"ur Festk\"orperforschung, Forschungszentrum J\"ulich, 
52425 J\"ulich, Germany}
%\e-mail{noguchi@issp.u-tokyo.ac.jp}
\begin{abstract}
Recent developments in lipid membrane models for simulations are reviewed.
To reduce computational costs, various coarse-grained molecular models
have been proposed. Among them, implicit solvent (solvent-free) molecular models are 
relatively more coarse-grained and efficient for simulating large bilayer membranes.
On a $\mu$m scale, the molecular details are typically negligible
and the membrane can be described as a continuous curved surface.
The theoretical models for fluid and elastic membranes
with mesh or meshless discretizations are presented.
As examples of applications, the dynamics of vesicles in flows,
vesicle formation, and membrane fusion are presented.
\end{abstract}
%\kword{bilayer membrane, vesicle, membrane fusion, red blood cell, solvent-free model, meshless method}

\maketitle

%\newpage
\section{Introduction}

An amphiphilic molecule, such as a lipid and detergent,
consists of hydrophilic (`water-loving') and hydrophobic (`water-fearing') parts.
In aqueous solution,
these molecules self-assemble into various structures depending on the relative
size of the hydrophilic part;
spherical or cylindrical-like micelles, bilayer membranes, and
inverted micelles.
In particular, the bilayer membrane of phospholipids is the basic structure
of the plasma membrane and intracellular compartments of living cells,
where the membranes are in a fluid phase and lipid molecules can diffuse 
in two-dimensional space.
A vesicle (closed membrane) is considered to be a simple model of cells
and also has applications as a drug-delivery system.

Many interesting behaviors of bilayer membranes have been studied such as
fluid-to-gel phase transition, lateral phase separation,
protein insertion, the lysis of lipid vesicles, and membrane fusion.
The thickness of a lipid membrane is $5$nm and cells are $\sim 10\mu$m in diameter.
Thus, the length scale of these phenomena varies from nm to $\mu$m.
To investigate the bilayer structure and molecular interactions,
information on a nm scale is necessary.
For example, the length mismatch of $1$nm between a protein and lipid 
can change the stability and interactions between proteins \cite{deme08}.
On the other hand, the morphologies of lipid vesicles and cells
on a $\mu$m scale can be understood by continuous surface theories \cite{safr94,lipo95,seif97}.

\begin{figure}[tb]
\begin{center}
\includegraphics{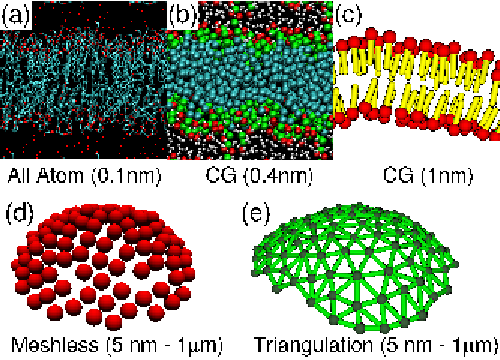}
\end{center}
\caption{(Color online)
Various membrane models. (a) all-atom model of C$_{12}$E$_2$.
(b) coarse-grained model with explicit solvent
constructed based on the all-atom simulation in (a)~\cite{shin08}.
(c) solvent-free molecular model~\cite{nogu01a}.
(d) meshless membrane model~\cite{nogu06}. (e) dynamically-triangulated membrane model~\cite{gg:gomp04c}.
Typical lengths of unit segments are shown in parentheses.
Figures (a) and (b) were provided by Shinoda.
}
\label{fig:cg}
\end{figure}

Various types of membrane models have been proposed for the simulations.
They are classified in two groups depending on whether the bilayer structure 
is implicitly or  explicitly taken into account.  
In the first group [Figs. \ref{fig:cg}(d), (e)],
the bilayer membrane is described as a smoothly-curved mathematical 
surface \cite{safr94,lipo95,seif97}.
This assumption is valid on length scales greater than the 
membrane thickness of $5$nm.  
Thus, a unit segment represents not a lipid molecule, but a membrane patch
consisting of hundreds to thousands of lipid molecules.
The information about the bilayer properties is only 
reflected in this case in the values of the elastic parameters.
A mesh discretization with a triangular or square mesh is typically employed 
for the simulation.
To allow large deformations of a fluid membrane, 
remeshing \cite{gg:gomp04c,gg:gomp97f,ma08} is necessary.
In particular, a dynamically-triangulated membrane model \cite{gg:gomp04c,gg:gomp97f} 
is widely used for the membrane with thermal fluctuations.
An alternative model is a meshless membrane,
in which particles self-assemble into a membrane by potential interactions.
The meshless models are very suitable for studying membrane dynamics accompanied 
by topological changes.

In the second group [Figs. \ref{fig:cg}(a)-(c)],
amphiphilic molecules are modeled by atomistic or
coarse-grained (CG) molecular models, 
and the solvent is taken into account either explicitly or implicitly.
Although computer technology is rapidly growing,
  $50$ ns dynamics of hundreds of lipid molecules
are typical levels
of recent simulations by the all-atom models.
Thus, CG models have been developed to reduce the computational costs.
CG molecular models with the Lenard-Jones potential \cite{goet98} and
with dissipative particle dynamics (DPD) \cite{hoog92,espa95} potential \cite{groo01}
have been widely used for the explicit-solvent simulations.
Recently, the potential parameters in the CG molecular models are tuned 
by atomistic simulations \cite{shel01,marr04,izve05,arkh08,shin08}, see Figs. \ref{fig:cg}(a) and (b).
To further reduce the computational costs,
larger CG segments (two or three segment particles per amphiphilic molecule)
are employed and the solvent is implicitly represented by
an effective attractive potential between the hydrophobic segments, see Figs. \ref{fig:cg}(c).
This type of implicit solvent model is typically called the solvent-free model.

There is a length-scale gap between these molecular models and the curved-surface models.
For the molecular models,
the size of the unit segments can be gradually varied
from $1\AA$ to $1$nm.
For the curved-surface models,
 a unit segment represents $\gtrsim 5$nm,
and the thermal undulations of smaller lengths are neglected.

Recently, the CG molecular simulations and triangulated membrane were reviewed
in Refs. \cite{niel04,muel06,vent06} and in Ref. \cite{gg:gomp04c}, respectively.
The phase separation in multicomponent membranes
was reported by Kumar in this JSPSJ special issue.
In this paper, we review the recent developments in 
the solvent-free molecular models and curved surface models
[Figs. \ref{fig:cg}(c)-(e)] for single-component membranes
and also describes some new results.
In Secs. \ref{sec:tri} and \ref{sec:meshless}, 
the triangulated and meshless membrane models are described, respectively.
The area-difference elasticity (ADE) model is newly applied 
to the dynamically-triangulated membrane.
The membrane dynamics in flows are presented for both models.
In Sec. \ref{sec:solf}, 
the solvent-free molecular models are described.
The membrane fusion and the formation of polyhedral vesicles
are presented as applications.

\section{Triangulated membrane} \label{sec:tri}

\subsection{fluid membrane}

\begin{figure}[tb]
\begin{center}
\includegraphics{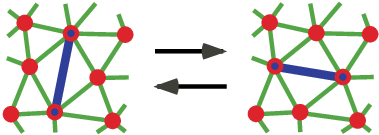}
\end{center}
\caption{(Color online)
Bond flip of triangulated mesh to produce membrane fluidity.
}
\label{fig:bflip}
\end{figure}

The  curvature energy of a fluid membrane is given by~\cite{safr94,canh70,helf73}
\begin{equation}
F_{\rm {cv}} =  \int  \Bigl[ \frac{\kappa}{2}(C_1+C_2 -C_0)^2  
                     + \bar{\kappa}C_1C_2 \Bigr]  dA
\label{eq:helf}
\end{equation}
where $C_1$ and $C_2$ are the principal curvatures at each point 
in the membrane.
The coefficients $\kappa$ and $\bar{\kappa}$
are the bending rigidity and saddle-splay modulus, respectively.
The spontaneous curvature $C_0$ vanishes when lipids symmetrically distribute 
in both monolayers of the bilayer.
Lipid membranes typically have $\kappa=20k_{\rm B}T$,
where $k_{\rm B}T$ is the thermal energy \cite{lipo95}.
For membranes of fixed topology without edges, the 
integral over the Gaussian curvature $C_1C_2$ is an invariant
and their properties are independent of $\bar{\kappa}$.
Since the critical micelle concentration (CMC) of lipids 
is very low, the number of lipid
molecules in a membrane is essentially constant over the typical experimental
time scales. Also, the osmotic pressure generated by 
ions or macromolecules in solution, which cannot penetrate the lipid
bilayer, keeps the internal volume essentially constant. 
Under the constraints of 
a constant volume $V$ and constant surface area $A$,
stomatocyte, discocyte, and prolate shapes
give the energy minimum of $F_{\rm {cv}}$ with $C_0=0$ for
$0<V^* \lesssim 0.59$, $0.59 \lesssim V^* \lesssim 0.65$, 
and $0.65 \lesssim V^* < 1$, respectively \cite{seif91},
where the reduced volume $V*= 3V/(4\pi R_0^3)$ with $R_0=(A/4\pi)^{1/2}$.

In a dynamically-triangulated surface model 
\cite{gg:gomp04c,gg:gomp97f,ho90,boal92,zhao05} of vesicles and 
cells, the membrane is described by $N$ vertices which are 
connected by bonds (tethers) to form a triangular network.
The vertices have excluded volumes.
The curvature energy is discretized 
using angles of the neighboring faces \cite{gg:gomp04c,gg:gomp97f,buen07} 
or using dual lattices of triangulation \cite{gg:gomp04c,gg:gomp97f,itzy86}.
To model the fluidity of the membrane, bonds
can be flipped to the diagonal of two adjacent triangles
using the Monte Carlo (MC) method, see Fig.~\ref{fig:bflip}. 
The membrane viscosity $\eta_{\rm {mb}}$ can be varied
by the frequency of the bond flip \cite{nogu04,nogu05}.
This model is widely used to simulate a vesicle with spherical topology.
The membrane with open 
edges  was also studied \cite{boal92,zhao05}.
Topological changes can be taken into account by 
a mesh reconnection \cite{gg:gomp04c,gomp98}. 
However, this reconnection is a discrete procedure,
so that it cannot smoothly treat the dynamics.

Recently, the dynamically-triangulated model
were applied to the dynamics of a fluid vesicle in flows
by combining with boundary element methods \cite{krau96,suku01}
or multi-particle collision (MPC) dynamics \cite{nogu04,nogu05,nogu07b,nogu05b}.
MPC is a particle-based simulation method proposed by Malevanets and Kapral \cite{male99}. 
In a simple shear flow, ${\bf v}=\dot\gamma y {\bf e}_x$,
the vesicle shows three types of motions \cite{krau96,nogu04,nogu05,nogu07b}.
(i) Tank-treading motion:
The vesicle has the constant inclination angle $\theta$ with respect to the flow direction,
and instead, the membrane is rotating.
(ii) Tumbling: The angle $\theta$ rotates.
(iii) Swinging: The angle $\theta$ oscillates around zero.
At a low shear rate $\dot\gamma$,
the vesicle transits from the tank-treading to tumbling with the increasing
viscosity contrast $\eta_{\rm {in}}/\eta_0$ or membrane viscosity $\eta_{\rm {mb}}$,
where $\eta_{\rm {in}}$ and $\eta_0$ are the viscosities of the internal and external fluids.
At the higher shear rate $\dot\gamma$,
a swinging phase appears between the tank-treading and tumbling phases.
These dynamics were also observed in experiments \cite{made06,kant06},
and are understood by the theory for an ellipsoidal vesicle \cite{kell82,nogu07b}
or the perturbation theory for a quasi-spherical vesicle \cite{seif99,misb06,lebe07}.
We found shear-induced shape transitions;
elongating transitions from stomatocyte and discocyte to prolate,
as well as a shrinking transition from a tumbling prolate to a tank-treading discocyte \cite{nogu04,nogu05}. 
In capillary flow, the fluid vesicle also transits from discocyte to prolate with the increasing flow velocity \cite{nogu05b}.

\subsection{area difference elasticity}

\begin{figure}[tb]
\begin{center}
\includegraphics{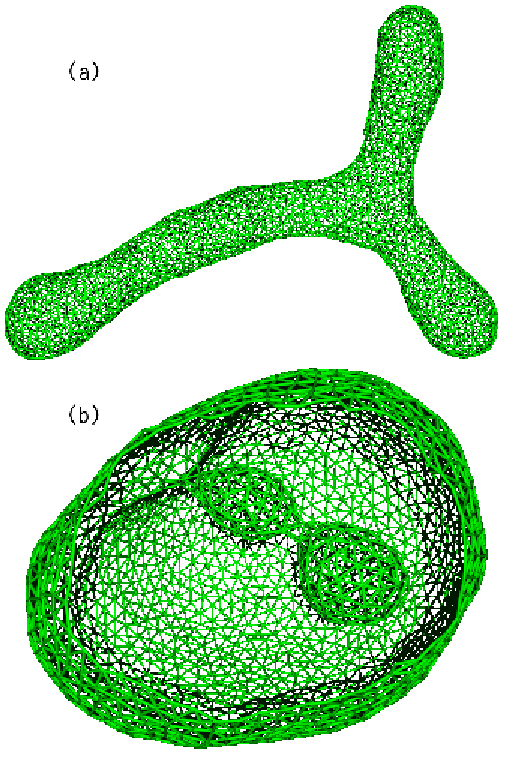}
\end{center}
\caption{(Color online)
Snapshots of a fluid vesicle of the ADE model
at $\kappa=k_{\rm {ade}}=20k_{\rm B}T$ and $N=2000$.
(a) Three-armed starfish-like shape at $V^*=0.4$ and $m_0^*=2$. 
(b) Two connected buds inside a vesicle at $V^*=0.8$ and $m_0^*=-0.5$.
The front quarter of the snapshot in (b) is removed to show the inside.
}
\label{fig:adesnap}
\end{figure}

In experiments, several morphologies of lipid vesicles
were observed in the addition to the above shapes: 
pear, peal-neckless, and branched starfish-like 
shapes \cite{lipo95,seif97,hota99,yana08}.
To explain them, the area difference elasticity (ADE) model
is widely used \cite{seif97,zihe05}.
In a vesicle, the area of two monolayers of the bilayer membrane
are different with  $\Delta A= h \oint (C_1+C_2) dA$,
where $h$ is the distance between two monolayers.
Since the flip-flop of lipids (traverse motion between monolayers) 
is very slow,
the area difference $\Delta A_0=(N_{\rm {out}}-N_{\rm {in}})a_0$
preferred by lipids is typically different from $\Delta A$,
where $N_{\rm {out}}$ and $N_{\rm {in}}$
are the number of lipids in the outer and inner monolayers, respectively,
and $a_0$ is the area per lipid.
In the ADE model, the energy of this mismatch $\Delta A-\Delta A_0$ is given by
\begin{equation}
F_{\rm {ade}} =  \frac{\pi k_{\rm {ade}}}{2Ah^2}(\Delta A - \Delta A_0)^2
=  \frac{k_{\rm {ade}}}{2}(m - m_0)^2,
\label{eq:ade}
\end{equation}
with the averaged curvature $m= (1/2R_0)\oint (C_1+C_2) dA$.
The curvature normalized by $m=4\pi$ of a spherical vesicle,
is denoted by $*$, as $m^*=m/4\pi$.
We simulated vesicles using the dynamically-triangulated model
for $F=F_{\rm {cv}}+F_{\rm {ade}}$ with
the harmonic constraint potentials of $A$ and $V$.
Figure \ref{fig:adesnap} shows snapshots of the Brownian dynamics (BD) simulation,
where the motions of the membrane vertices are given by the underdamped Langevin equations.

\begin{figure}[tb]
\begin{center}
\includegraphics{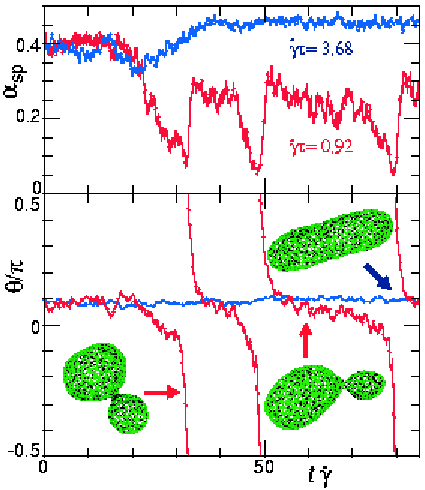}
\end{center}
\caption{(Color online)
Time developments of asphericity $\alpha_{\rm {sp}}$ and inclination angle $\theta$
of a vesicle in simple shear flow at $V^*=0.75$, $\kappa=20k_{\rm B}T$,
 $k_{\rm {ade}}=14k_{\rm B}T$, $m_0^*=2$, $\eta_{\rm {in}}/\eta_0=1$, 
$\eta_{\rm {mb}}=0$, and $N=500$.
The vesicle exhibits shape transitions from a budded shape to prolate at $\dot\gamma=3.68$
and from prolate to the budded shape at $\dot\gamma=0.92$.
The asphericity is the degree of deviation from a spherical shape \cite{rudn86},
$\alpha_{\rm {sp}} = \{({\lambda_1}-{\lambda_2})^2 + 
  ({\lambda_2}-{\lambda_3})^2+({\lambda_3}-{\lambda_1})^2\}/(2 R_{\text g}^4)$,
where ${\lambda_1} \leq {\lambda_2} \leq {\lambda_3}$ are the 
eigenvalues of the gyration tensor of the vesicle and
$R_{\rm g}^2=\lambda_1+\lambda_2+\lambda_3$.
The intrinsic time unit is $\tau=\eta_0 R_0^3/\kappa$. 
The other simulation parameters are described in Ref. \cite{nogu05}.
}
\label{fig:adeshear}
\end{figure}

Vesicles of the ADE model also exhibit shape transitions in shear flow.
A budded vesicle [see the bottom snapshots in Fig. \ref{fig:adeshear}]
shows a tumbling motion at the low shear rate $\dot\gamma=0.92$
[see red lines in Fig. \ref{fig:adeshear}],
even for $\eta_{\rm {in}}/\eta_0=1$ and $\eta_{\rm {mb}}=0$,
where prolate and oblate vesicles at any $V^*$ only show tank-treading.
A similar tumbling motion was reported for a budded  two-component vesicle \cite{smit07}.
At the higher shear rate $\dot\gamma=3.68$, 
the vesicle is found to transit into a prolate shape,
which shows tank-treading [see blue lines in Fig. \ref{fig:adeshear}].
As the shear rate decreases, the vesicle transits back to a budded shape.
The vesicle dynamics of other shapes, such as starfishes, 
are an interesting problem for further studies.

\begin{figure}[tb]
\begin{center}
\includegraphics{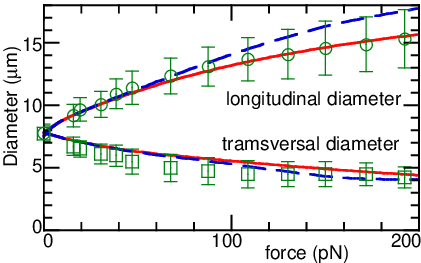}
\end{center}
\caption{(Color online)
Diameters of an RBC stretched by optical tweezers.
The symbols represent the experimental data from Ref. \cite{mill04}.
The broken and solid lines represent the simulation data at $k_2=0$ and $1$,
respectively}
\label{fig:rbcstr}
\end{figure}

\begin{figure}[tb]
\begin{center}
\includegraphics{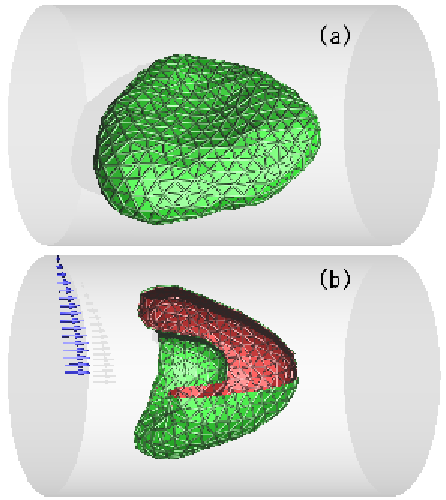}
\end{center}
\caption{(Color online)
Snapshots of an elastic vesicle in capillary flow at $\kappa=20k_{\rm B}T$ and 
$\mu R_{\rm 0}^2/k_{\rm B}T=110$ \cite{nogu05b}.
(a) Discocyte shape in slow flow. (b) Parachute shape in fast flow. 
The upper-front quarter of the snapshot in (b) is removed to show the inside.
The arrows represent the flow velocities.}
\label{fig:rbccap}
\end{figure}

\subsection{elastic membrane}

The membranes (shells) of synthetic capsules, virus, and red blood cells (RBCs),
have the shear elasticity $\mu$. We call them elastic membranes.
The elastic membrane can be modeled as a fixed mesh (network).
The relative importance of the bending and shear elasticities is determined by the 
dimensionless F\"oppl-von K\'arm\'an number
$\gamma=E R_0^2/\kappa$, where $E$ is the 
two-dimensional Young modulus \cite{lidm03}. 
The elastic membrane with $12$ disclinations
has a spherical shape at $\gamma \lesssim 150$
and an icosahedral shape at $\gamma \gtrsim 200$.
The bending or shear elasticity is dominant for
the former or latter condition, respectively.
The properties of viral capsids are well explained by 
the elastic membrane models \cite{lidm03,vlie06,buen07}. 

The models of an RBC membrane have been intensively studied 
\cite{skal73,disc98,nogu05b,dao06,dupi07,vazi08,pivk08}.
In the molecular-scale models \cite{disc98,dao06}, 
the spectorin network attached on the bilayer membrane 
is modeled by bonds with a worm-like chain (WLC) potential.
In the continuum models,
the Skalak model \cite{skal73} is widely used.
Recently, the comparison with the optical tweezer experiment \cite{mill04}
became a standard method to tune the elastic parameters \cite{dao06,dupi07,vazi08,pivk08}.
Figure \ref{fig:rbcstr} shows the experimental data \cite{mill04} and our simulation data.
In the simulation, the RBC membrane is modeled by triangular networks with $578$ vertices,
which are connected by a bond potential $U_{\rm {bond}}=(k_1/2)(r-r_0)^2\{1+ (k_2/2)(r/r_0-1)^2\}$
with $\mu=(\sqrt{3}/4)k_1=6\times 10^{-6}N/m$ and $\kappa=2 \times 10^{-19}$J.
The nonlinear term with $k_2=1$ gives better agreement with the experimental data,
but an arbitrariness in the choice of the nonlinear function  still remains.
The shape transitions of RBC with a stomatocyte-discocyte-echinocyte sequence can be 
reproduced by the elastic membrane with the ADE model \cite{lim02}.

The dynamics of the RBCs and elastic capsules have been simulated 
by the elastic membrane combined with boundary element methods \cite{pozr01,pozr05b,lac08},
immersed boundary methods \cite{eggl98,liu06,dupi07,sui08}, MPC \cite{nogu05b}, and DPD \cite{pivk08}.
Figure \ref{fig:rbccap} shows the elastic vesicles in capillary flow \cite{nogu05b}.
At the small flow velocities, the symmetry axis of the 
discocyte is found not to be oriented perpendicular to the 
cylinder axis. With the increasing flow velocity, the elastic vesicle
transits into a parachute-like shape
while the fluid vesicle transits into a prolate shape.
The transition velocity linearly depends on the elasticities $\mu$ and $\kappa$.
The results are in good agreement with the experimental data \cite{suzu96}.
In simple shear flow, 
elastic capsules \cite{sui08,kess08,chan93} and RBCs \cite{pivk08,abka07}
 transit from tumbling to tank-treading with shape oscillation (swinging)
as the shear rate increases.
These dynamics are caused by an energy 
barrier for the tank-treading rotation \cite{skot07}.

\begin{figure}[tb]
\begin{center}
\includegraphics{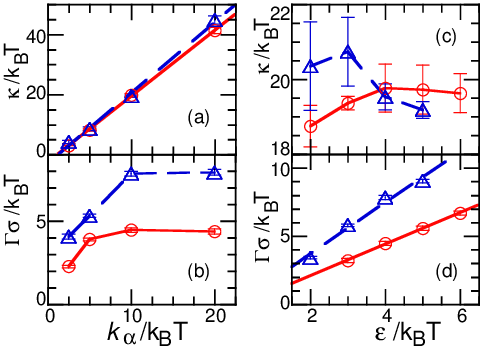}
\end{center}
\caption{(Color online)
Parameter $k_{\alpha}/k_{\rm B}T$ and $\varepsilon$ dependence of 
 the bending rigidity $\kappa$ and the line tension $\Gamma$ \cite{nogu06}.
The solid (red) lines represent data for $r_{\rm {att}}/\sigma=1.8$ and $\rho^*=6$.
The dashed (blue) lines represent data for $r_{\rm {att}}/\sigma=1.9$ and $\rho^*=8$.
The solid and dashed lines in (a) and (c) show linear fits.
}
\label{fig:mlspara}
\end{figure}

\begin{figure}[tb]
\begin{center}
\includegraphics[width=8.5cm]{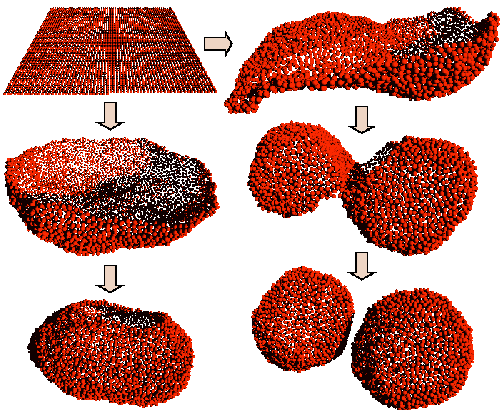}
\end{center}
\caption{(Color online)
Sequential snapshots of membrane closure simulated by 
the meshless membrane model with BD at $N=4000 \simeq 1000 N_{\rm 0}$, 
$r_{\rm {att}}/\sigma=1.8$, $\rho^*=6$, $k_{\alpha}/k_{\rm B}T=5$, 
and $\varepsilon/k_{\rm B}T=6$ \cite{nogu06a}.
The membrane stochastically forms one or two vesicles.
}
\label{fig:mlsclo}
\end{figure}

\begin{figure}[tb]
\begin{center}
\includegraphics{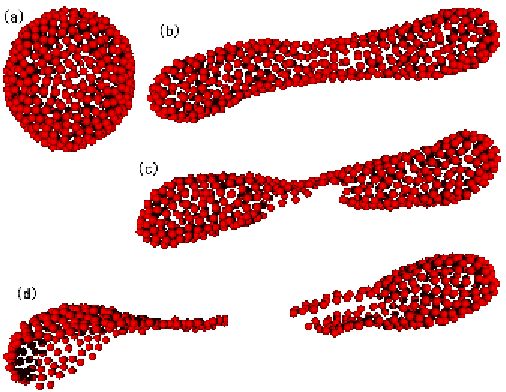}
\end{center}
\caption{(Color online)
Sequential snapshots of a vesicle of the meshless membrane
 in simple shear flow simulated by MPC at $\dot\gamma = 3.7 \kappa/(\eta R_{\rm {int}}^3)$,
where $R_{\rm {int}}=7.6\sigma$ is the radius of the initial vesicle of $500$ particles.
The membrane is in a fluid state with
 $r_{\rm {att}}/\sigma=1.8$, $\rho^*=6$, $k_{\alpha}/k_{\rm B}T=10$, 
and $\varepsilon/k_{\rm B}T=3$.
(a) $t=0$, (b) $t=15/\dot\gamma$, (c) $t=18/\dot\gamma$, and (d) $t=22/\dot\gamma$.}
\label{fig:mlsshear}
\end{figure}

\section{meshless membrane} \label{sec:meshless}

The meshless models are particle-based methods,
which do not need bond connections between neighboring particles.
All of the proposed meshless methods are solvent-fee methods, 
but can be extended to explicit solvent versions.

The first meshless model was proposed  by Drouffe {\it et al.} 
in 1991 \cite{drou91}. 
The particles possess an orientational degree of freedom and interact 
with each other via three potentials: a soft-core repulsion, 
an anisotropic attraction, and a hydrophobic multibody interaction. 
The particles self-assemble into a fluid membrane.
Very recently, the modified models with pairwise potentials
were proposed \cite{popo08,kohy09}. 
They also have orientational degrees of freedom 
and can form fluid membranes.

Recently, we proposed an alternative meshless model 
\cite{nogu06} in which particles possess no internal degrees of 
freedom --- unlike the other models.
In our model, the particles interact with each other via the potential
\begin{equation} \label{eq:Umls}
U= \varepsilon( U_{\rm {rep}} +  U_{\rm {att}}) 
                      + k_{\rm {\alpha}} U_{\rm {\alpha}},
\end{equation}
which consists of a repulsive soft-core potential $U_{\rm {rep}}$ with a 
diameter $\sigma$, an attractive potential $U_{\rm {att}}$, and a 
curvature potential $U_{\alpha}$.  All three potentials only depend on 
the positions ${\bf r}_i$ of the particles.
Two types of curvature potentials \cite{nogu06} were proposed on the basis of
the moving least-squares (MLS) method~\cite{bely96,lanc81}.
Here, we only describe the simpler potential,
which drives the particles onto a plane: 
$U_{\rm {\alpha}}= \sum_i \alpha_{\rm {pl}}({\bf r}_{i})$.
The degree of deviation from a plane, the aplanarity $\alpha_{\rm {pl}}$,
is defined as
\begin{equation}\label{eq:alpl}
\alpha_{\rm {pl}}
 = \frac{9\lambda_1\lambda_2\lambda_3} {(\lambda_1+\lambda_2+\lambda_3)
    (\lambda_1\lambda_2+\lambda_2\lambda_3+\lambda_3\lambda_1)},
\end{equation}
where ${\lambda_1}$, ${\lambda_2}$, and ${\lambda_3}$ are
 eigenvalues of the weighted gyration tensor,
$a_{\alpha\beta}= \sum_j (\alpha_{j}-\alpha_{\rm G})
(\beta_{j}-\beta_{\rm G})w_{\rm {mls}}(r_{i,j})$
with $\alpha, \beta=x,y,z$ and a smoothly-truncated 
Gaussian function $w_{\rm {mls}}(r)$ \cite{nogu06}.
In this model, $\kappa$ and $\Gamma$ can essentially
be independently varied, see Fig. \ref{fig:mlspara}.
The bending rigidity $\kappa$ linearly increases with $k_{\rm {\alpha}}$ 
and is almost independent of $\varepsilon$.
On the other hand, the line tension $\Gamma$ linearly increases with 
$\varepsilon$ and is
almost independent of $k_{\rm {\alpha}}$ for $k_{\alpha}/k_{\rm B}T \gtrsim 10$.
The saddle-splay modulus is roughly estimated as
$\bar{\kappa}\simeq -\kappa$. 

Although the hydrodynamic interactions are not present in the model itself,
 these interactions can be taken into account by combining 
 with a particle-based hydrodynamic method
 such as MPC \cite{male99} and DPD \cite{hoog92,espa95}.
Thus, the hydrodynamic interactions can be easily switched on or off.
Static equilibrium properties can be investigated by  
BD or MC, which requires much less computational time.
One of the disadvantages of meshless methods,
compared with the mesh models,
is that the vesicle volume cannot be constrained,
since the exact volume is difficult to calculate.
However, the volume can be naturally fixed
if an explicit solvent is introduced.
 
The effects of the hydrodynamic interactions for the self-assembly and dissolution
 are investigated by comparing MPC with BD \cite{nogu06a}.
The hydrodynamic interactions are found to speed up the 
dynamics in both cases.  
The particles self-assemble into discoidal clusters, and
clusters aggregate, and then
large clusters form vesicles.
To quantitatively investigate the closing dynamics to a vesicle,
the simulations of the initially flat membranes were performed.
For a membrane slightly above the critical size $4N_{\rm {0}}$,
all of them close into vesicles via a bowl-like shape.
At $N=N_{\rm {0}}$,
the line tension energy of a flat disk equals the bending energy 
of a spherical vesicle \cite{from83}.
For much larger membranes of $N \gg 4N_{\rm {0}}$,
however, the membrane stochastically forms two vesicles
via an $S$-shaped conformation, see Fig. \ref{fig:mlsclo}.
The large line tension induces a negative surface tension,
and the resulting buckled shape can grow.
The closing vesicle has an oblate shape in the BD simulations with large $N$
[see the left-bottom snapshot in Fig. \ref{fig:mlsclo}].
With the hydrodynamic interactions,
the pressure of the internal fluid makes this closing shape more 
spherical.

When the line tension is reduced,
a vesicle dissolves via an opened pore on the membrane
 with $\sqrt{N(t)}= \sqrt{N(0)} - c t/2$ for both MPC and BD \cite{nogu06a}.
A similar pore-opening was observed in the lysis of a lipid vesicle 
by detergents \cite{nomu01}.

In simple shear flow, a spherical vesicle of the meshless membrane
elongates into a prolate shape with its decreasing volume.
Similar elongation is observed for the vesicles consisting of surfactants \cite{shah98}.
At higher shears, the vesicle ruptures into two pieces, see Fig. \ref{fig:mlsshear}.
The membrane rupture is a fundamental process for
the formation of multi-lamellar vesicles in shear flows \cite{diat93b,mort01,medr05}.

\begin{figure}[tb]
\begin{center}
\includegraphics{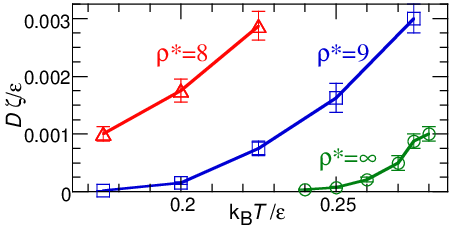}
\end{center}
\caption{(Color online)
Temperature $T$ dependence of diffusion constant $D$
of amphiphilic molecules, consisting of one hydrophilic segment and one hydrophobic segment 
in the bilayer membrane for various cutoff densities $\rho^*$.
}
\label{fig:dif}
\end{figure}

\section{Solvent-free molecular model}\label{sec:solf}
\subsection{model}

The simulations of lipid molecules with an explicit solvent require 
the calculation for a large number of water molecules 
in addition to lipid molecules.  
A small patch of flat membranes
can be simulated with $30$ water molecules per lipid by the atomistic models \cite{tiel97}.
However, more water molecules are needed for the simulations of  
vesicles, since the formation of a vesicle (Fig.~\ref{fig:mlsclo}) 
requires large solvent space to prevent membrane interactions 
through the periodic boundary conditions of the simulation box.
The self-assembly of amphiphilic molecules in dilute solutions
requires many more water molecules.
The solvent-free models are more efficient tools for the simulations
which require a larger solvent space. A similar solvent-free approach 
is also frequently used in the simulations of polymers and proteins.

The first solvent-free molecular model, 
which self-assembles into a bilayer membrane without an explicit solvent,
was proposed by us in $2001$ \cite{nogu01a}.
Later, the modified models
were proposed for membranes \cite{fara03,bran04,bran05,cook05,wang05,lyub05,reva08}  and for micelles \cite{mori05}.
An amphiphilic molecule is modeled as rigid \cite{nogu01a,fara03,bran04} 
or flexible \cite{bran05,cook05,wang05,lyub05,reva08,mori05} chains
with single \cite{fara03,bran04,bran05,cook05,wang05,reva08,mori05} 
or double \cite{lyub05} hydrophobic tails.
The molecules interact with each other with
pairwise \cite{fara03,bran04,bran05,cook05,lyub05,reva08} or multibody 
\cite{nogu01a,wang05,mori05} potentials.
In particular, the potentials in Refs. \cite{lyub05,mori05}
are constructed from the radial distribution function of atomistic simulations.

A common feature of the models is 
the requirement of an attractive potential between
hydrophobic segments.
This attraction mimics the ``hydrophobic'' interaction
(hydrophobic segments dislike to contact with water).
One of the simple ideas to estimate this interaction
is that the hydration energy is assumed to be proportional to 
the solvent accessible surface area (ASA) \cite{feig04,ooi87}.
Since the calculation of the ASA is a numerically time-consuming task,
an effective potential, which is a function of the local density $\rho$ of hydrophobic 
segments, were proposed for protein \cite{taka99} and lipid molecules \cite{nogu01a}.
At a low density $\rho<\rho^*-1$, the potential acts as a pairwise attractive potential, but
at a high density $\rho>\rho^*$, this attraction vanishes, in which the segments 
are assumed to be completely surrounded 
by other hydrophobic segments.
This multibody potential can produce
a very fast lateral diffusion of molecules
and a wide fluid-phase range ($0.1\lesssim k_{\rm B}T/\varepsilon \lesssim 0.9$) \cite{nogu01a}.
A fluid membrane is also generated by pairwise attractive potentials,
but their CMC is very high,
and the membrane often coexists with isolated monomers.
Although the range of the fluid phase becomes larger for wider pairwise 
potentials \cite{cook05},
their ranges are still relatively small.
Figure \ref{fig:dif} shows that the lateral diffusion on the membrane becomes faster
with the decreasing $\rho^*$, where $\rho^*=\infty$ corresponds to a pairwise potential.
The attractive pairwise potential of other solvent-free models can be
 extended to density-dependent potentials in order to obtain faster dynamics.

\begin{figure}[tb]
\begin{center}
\includegraphics{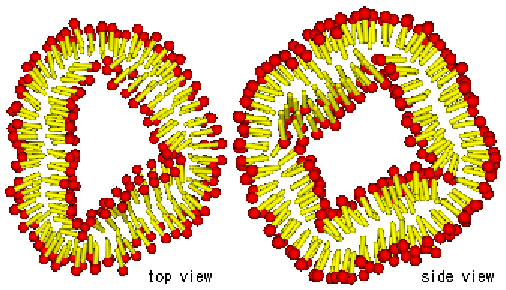}
\end{center}
\caption{(Color online)
Sliced snapshots of a triangular-prism-shaped vesicle
at $N=1000$, $k_{\rm B}T/\varepsilon=0.2$ and $C_0\sigma=0.11$ \cite{nogu03}.
The red spheres and yellow cylinders
represent the hydrophilic and hydrophobic segments of amphiphilic
molecules, respectively.
}
\label{fig:hedo}
\end{figure}

The solvent-free molecular models have been applied to a variety of phenomena:
self-assembly to vesicles \cite{nogu01a,lyub05},
membrane fusion \cite{nogu01b,nogu02a,nogu02c}, 
membrane fission \cite{nogu02a,nogu02b,nogu03}, 
the formation of polyhedral vesicles \cite{nogu03},
pore formation \cite{fara03,ilya08},
the adhesion of nanoparticles \cite{nogu02a,reyn07},
the fluid-gel phase transition \cite{bran06,reva08}, 
phase separation of lipids \cite{cook05}, 
protein inclusion in membrane \cite{bran06,ilya08}, and
DNA-membrane complexes \cite{fara06}.
Figure \ref{fig:hedo} shows a polyhedral vesicle.
When the bending rigidity is large compared with the vesicle radius,
the lined defects at the edges of polyhedrons are induced by
the mismatch of the membrane curvature with the spontaneous curvature $C_0$ of the monolayer.
The number of edges and vertices increases with the increasing $C_0$ \cite{nogu03}.
Recently, a similar defect was proposed as a kink structure in a symmetric ripple phase \cite{lenz07}.

In early years,
an alternative implicit-solvent molecular model, in which
the head segment is constrained on a plane,
was employed \cite{ploe82,sint98}.
In this model, the membrane cannot be deformed or self-assembled.
However, the diffusive motion of molecules is still present on a fixed membrane geometry
and it may provide a good reference state to investigate the effects of
the thermal undulations of a membrane.
Recently, a phantom-solvent model was proposed \cite{lenz05,lenz07},
where an explicit but very simple solvent is employed.
In this model, solvent particles have a repulsive interaction with amphiphilic molecules,
but do not interact with each other.
Thus, the solvent is in the ideal-gas equation of state, $PV=nk_{\rm B}T$, so that
its pressure is easily controlled.

\subsection{membrane fusion and fission}

\begin{figure}[tb]
\begin{center}
\includegraphics{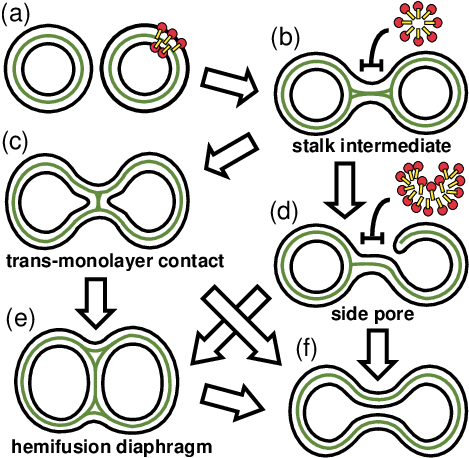}
\end{center}
\caption{(Color online)
Schematic representation of the fusion pathways obtained by molecular simulations.
 The green (gray) lines represent the hydrophobic boundaries of two monolayers.
}
\label{fig:fuspath}
\end{figure}

The membrane fusion and fission are  key events in various intra- and intercellular
processes, such as protein trafficking, fertilization, and
viral infection.
In an endocytosis pathway, 
a small vesicle pinches off from the plasma membrane and fuses
with a lysosome.
Several fusion mechanisms and fusion intermediate structures have been proposed
for the fusion of biological and lipid membranes~\cite{jahn02}. Among them,
the stalk model~\cite{muel06,jahn02,cher08,mark84,sieg93,kozl02,mark02}
 is widely accepted and is qualitatively supported by
experimental studies.
The first intermediate in this model, a stalk, is a
hourglass-like structure that connects only the outer monolayers of
the vesicles [Fig. \ref{fig:fuspath}(b)]. 
In the stalk models,
two types of pore-opening pathways from the stalk intermediate
were previously proposed.
Radial expansion of the stalk results in contact between
the inner monolayers inside the stalk, a trans-monolayer contact state
 [Fig. \ref{fig:fuspath}(c)]. 
In the original model~\cite{mark84}, 
expansion of the contact area results in a disk-shaped bilayer consisting of both inner
monolayers, called a hemifusion diaphragm [Fig. \ref{fig:fuspath}(e)], and a fusion pore is formed in
the hemifusion diaphragm [Fig. \ref{fig:fuspath}:(b)$\rightarrow$(c)$\rightarrow$(e)$\rightarrow$(f)]. 
The original model is based only on the calculation of the bending energy of
the monolayers in fusion intermediates.
 Siegel~\cite{sieg93} also took into account the
interstice (void) energy of junctions of monolayers.
In the model modified by Siegel, the trans-monolayer contact results in pore formation,
rather than expansion of the contact area [Fig. \ref{fig:fuspath}:(b)$\rightarrow$(c)$ \rightarrow$(f)].
Although the free energy barrier was estimated to be too high ($\sim200k_{\text{B}}T$) in Ref. \cite{sieg93},
the barrier is reduced by a more quantitative estimation of the monolayer geometry \cite{cher08,mark02}.
The molecular tilt in the monolayers was proposed as the mechanism to fill the void space \cite{kozl02}.
Recently, the free energy landscape of the fusion was also studied
using the self-consistent field theory \cite{muel06,lee08,sevi05}.

\begin{figure}[tb]
\begin{center}
\includegraphics{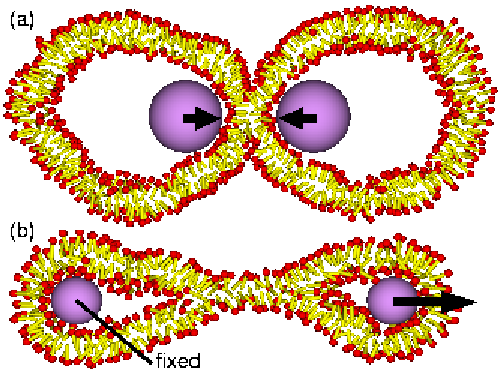}
\end{center}
\caption{(Color online)
Sliced snapshots of vesicles under external forces.
(a) Trans-monolayer contact in the membrane fusion \cite{nogu02c}.
(b) Formation of a stalk (cylindrical) structure in
the stretched vesicle \cite{nogu02b}. 
The violet spheres represent nanoparticles. 
}
\label{fig:fussnap}
\end{figure}

The first molecular simulation of membrane fusion was performed by a solvent-free model \cite{nogu01b}.
Small vesicles spontaneously fuse via the modified stalk pathway 
[Fig. \ref{fig:fuspath}:(b)$\rightarrow$(c)$ \rightarrow$(f)] at low temperature. 
However, at high temperature, the vesicles fuse via a new pathway,
where a pore opens on the side of the stalk
and then the elongated stalk bends to form  a fusion pore 
[Fig. \ref{fig:fuspath}:(b)$\rightarrow$(d)$ \rightarrow$(f)].
The adhesion of a nanoparticle was found to promote this fusion process \cite{nogu02a}.
Recently, the leakage between the interior and exterior of a vesicle 
was observed in an experiment \cite{frol03}.
This leakage supports the side-pore pathway.
The vesicles pinched by particles form the trans-monolayer contact [Fig. \ref{fig:fussnap}(a)]
and then fusion directly occurs [Fig. \ref{fig:fuspath}:((c)$ \rightarrow$(f)] or via 
the hemifusion diaphragm [Fig. \ref{fig:fuspath}:(c)$\rightarrow$(e)$ \rightarrow$(f)]  \cite{nogu02c}.

Membrane fission occurs via pathways opposite of the fusion.
In a stretched vesicle, a stalk structure [Fig. \ref{fig:fussnap}(b)] is formed via the trans-monolayer contact 
[Fig. \ref{fig:fuspath}:(f)$\rightarrow$(c)$ \rightarrow$(b)] \cite{nogu02b}.
On the other hand, the particle adhesion induces pore opening on the necked membrane and
the pore expansion leads to the stalk formation  [Fig. \ref{fig:fuspath}:(f)$\rightarrow$(d)$ \rightarrow$(b)] \cite{nogu02a}.
In large spontaneous curvature of the monolayers induces the vesicle fission via pore-opening at the lined defects \cite{nogu03}.

Recently, membrane fusion was simulated by various molecular models such as 
the bond-fluctuation lattice model \cite{muel02,muel03}, explicit-solvent CG models \cite{marr03,stev03,smei06,kass07},
DPD models \cite{li05,shil05,gao08}, and the atomistic model \cite{knec07}.
These simulations also show fusion pathways as shown in Fig. \ref{fig:fuspath}.
The hemifusion diaphragm was found to also be formed from the side-pore of the stalk \cite{marr03,smei06,gao08}.
The membrane fission was also simulated in two-component vesicles \cite{smit07,yama03,mark07}.
Although the dependence on external forces \cite{nogu02c} and surface tension \cite{muel06,muel03,shil05,gao08}
was investigated, conditions to determine the pathways are not fully understood.
The membrane fusion and fission are induced by proteins in living cells.
Their molecular mechanisms are not well understood and have not yet been simulated.

\section{Summary}

We have presented three types of membrane models: the triangulated membrane, 
meshless membrane, and solvent-free molecular models.
The first two models are constructed for large scale ($\mu$m)
and the last model is for a molecular scale (nm).
To study the bending deformation of membranes, 
the details on a molecular scale are typically not necessary,
therefore, the mesh and meshless models are suitable and much more efficient than the molecular models.
On the other hand, 
to study the phenomena including the non-bilayer structure,
such as membrane fusion and protein insertion, 
the molecular scale cannot be neglected.

When compared with the phenomena in thermal equilibrium,
the nonequilibrium phenomena are not well understood.
In this paper, we presented several dynamic behaviors on a $\mu$m-scale in flows.
On the other hand, 
the dynamics on a molecular scale were much less explored while
a few simulation studies \cite{shku05,guo07} were reported.
The electric field is another interesting external field,
which can open pores on a membrane \cite{boch08}
and induce shape deformation of a vesicle \cite{risk06}.
In living cells, the biomembrane is in a complex nonequilibrium environment.
The membrane models presented in this paper are powerful tools to study the membrane behaviors
under nonequilibrium as well as under equilibrium conditions. 

\section*{Acknowledgment}
The author would like to thank W. Shinoda (AIST) for providing the figures,
and G. Gompper (J{\"u}lich) for the helpful discussions. 

%\bibliography{mls,tri}

{\bf Hiroshi Noguchi} was born in Hyogo prefecture, Japan, in 1973. 
He received his B. Sc. (1995), M. Ph. (1997), and D. Ph. (2000) from Nagoya University.
He was a postdoc at the Institute for Molecular Science (Okazaki) (2000-2003),
and a postdoc (2003-2006) and a research staff member (2006-2008) at Forschungszentrum J\"ulich.
He has been an associate professor at the University of Tokyo since May 2008.
He has studied soft matter physics, in particular, membrane physics, using theories and simulations.

\end{document}